\documentclass[a4paper]{jpconf}

\usepackage{graphicx}
\usepackage{amsmath,amssymb}

\newcommand{\ii}{\mathrm{i}}
\newcommand{\dd}{\mathrm{d}}
\newcommand{\dc}{\delta_{\cal C}}
\newcommand{\thc}{\theta_{\cal C}}
\newcommand{\pdert}{\partial_{t}}
\newcommand{\tc}{T_{\cal C}}
\newcommand{\intc}{\int_{\cal C}}
\newcommand{\aux}{\mathrm{aux}}

\begin{document}


\title{The auxiliary Hamiltonian approach and its generalization to non-local self-energies}
\author{Karsten Balzer}
\address{Rechenzentrum, Christian-Albrechts-Universit\"{a}t zu Kiel, Ludewig-Meyn-Strasse 4, 24118 Kiel, Germany}
\ead{balzer@rz.uni-kiel.de}


\begin{abstract}
The recently introduced auxiliary Hamiltonian approach [Balzer K and Eckstein M 2014 \textit{Phys.~Rev.~B}
\textbf{89} 035148] maps the problem of solving the two-time Kadanoff-Baym equations onto a noninteracting
auxiliary system with additional bath degrees of freedom. While the original paper restricts the discussion
to spatially local self-energies, we show that there exists a rather straightforward generalization to
treat also non-local correlation effects. The only drawback is the loss of time causality due to a combined
singular value and eigen decomposition of the two-time self-energy, the application of which inhibits one to
establish the self-consistency directly on the time step. For derivation and illustration of the method, we
consider the Hubbard model in one dimension and study the decay of the N\'{e}el state in the weak-coupling
regime, using the local and non-local second-order Born approximation.
\end{abstract}


\section{Introduction}\label{sec.sec1}
Many numerical approaches which are based on the Keldysh formalism require the solution of the two-time
Kadanoff-Baym equations (or the nonequilibrium Dyson equation) which are equations of motions for the
one-particle nonequilibrium Green's function defined on the \mbox{L-shaped} Keldysh contour~\cite{kadanoff62,keldysh64,stefanucci13.cup}.
The particular double-time and non-Markovian form of these integro-differential equations generally renders
their  integration a non-trivial and challenging task. Moreover, big efforts are needed to reach either
long times, handling a large memory kernel, or to simulate quantum systems with many degrees of freedom
and especially spatially inhomogeneous systems.

Despite these obstacles, much computational progress has been made over the recent decades, driven by 
the widespread success of nonequilibrium Green's functions in various fields, including nuclear~\cite{danielewicz84.a,danielewicz84.b,kohler01},
plasma~\cite{kwong00,kremp05}, semiconductor~\cite{kwong98,gartner99,lorke06}, condensed matter~\cite{freericks06,aoki14}
and atomic and molecular physics~\cite{dahlen07,perfetto15}. Following the pioneering work of Refs.~\cite{danielewicz84.a,danielewicz84.b},
which focuses on the numerical simulation of heavy-ion collisions, time propagation schemes of the
Kadanoff-Baym equations (KBE) have been developed in different contexts to study the correlation build-up
in homogeneous fermion systems~\cite{morawetz99,kremp00}. A general Fortran code is presented in Ref.~\cite{kohler99.fortran}
which efficiently evaluates the collision integrals (convolutions) by using fast Fourier transforms of the
momentum resolved Green's function. Concerning the treatment of initial correlations different pathways
have been pursued which range from the use of adiabatic switching~\cite{rios11,hermanns12.gkba} and the
embedding of additional collision terms derived from the equilibrium pair correlation function~\cite{semkat99}
to the extension of the real-time Green's function onto the imaginary branch of the Keldysh contour~\cite{morozov99,dahlen05}.
With the advance of computer power, also finite and inhomogeneous systems have become accessible. A corresponding
well recognized predictor-corrector-based time propagation code is described in Ref.~\cite{stan09} and is
used in many applications. Further progress is due to adapted basis sets which have been deployed to simplify
the interaction (Coulomb) matrix elements, e.g., \cite{balzer10.pra1}, and due to efficient parallelization
strategies~\cite{garny10,balzer10.pra2,balzer13.lnp}, where the Green's function is distributed over the
memory of multiple compute nodes in such a way that the collision integrals can be calculated locally and
a minimum communication is required to perform the time stepping.

A feature, that is common to virtually all time propagation codes, is the direct evaluation of the memory
kernel, i.e., the computation of the collision integrals which cover the whole time history. Together with
the double-time structure of the Green's function, this leads to the typical $n_t^3$-scaling of the numerical
algorithms in the number of time steps. Only further approximations can overcome this unfavorable scaling,
such as the application of the generalized Kadanoff-Baym ansatz~\cite{hermanns12.gkba,lipavsky86,hermanns13.jpcs},
which reconstructs the two-time Green's function from its time-diagonal value and is expected to be a reasonable
option to extend state-of-the art simulations, e.g.,~\cite{latini14,balzer13.jpcs,hermanns14.prb}.

The prevailing similarity of the time propagation schemes raises the question whether there exist alternative
methods to solve the two-time KBE. Aside from matrix inversion techniques, e.g.~\cite{freericks08}, a promising
idea has been triggered by nonequilibrium dynamical mean-field theory (DMFT)~\cite{aoki14}, where Hamiltonian-based
impurity solvers have recently become available through a two-time decomposition of the hybridization
function~\cite{gramsch13,wolf14.mps,balzer15.mctdh,balzer15.neel}. In Ref.~\cite{balzer14.aux} it has been
shown that a similar decomposition of the self-energy can be used to derive an auxiliary Hamiltonian representation
of the Kadanoff-Baym equations, where the one-particle nonequilibrium Green's function of the interacting many-body
problem under consideration is obtained from a \textit{noninteracting} auxiliary system which couples to additional
bath orbitals the number of which is finite. Although progress has been made recently by showing that a nonequilibrium
self-energy has a unique Lehmann representation~\cite{gramsch15}, the auxiliary Hamiltonian method has so far been
proven to be effective only for a local self-energy, which represents an approximation becoming exact in the limit of
infinite spatial dimensions (and is just the central approximation in DMFT).

In this contribution, we show that the method can be extended rather easily to treat also non-local self-energies
and that it is thus applicable more generally. Particularly, it can improve the $n_t^3$-scaling in the same manner
as in Ref.~\cite{balzer14.aux} as long as the computation of the self-energy itself represents a sufficiently simple task.
Note that this is a rather general prerequisite of the approach, independent of the self-energy's spatial form.
We start the discussion with a brief review of the
auxiliary Hamiltonian approach for systems with a local self-energy (Sec.~\ref{subsec.subsec2.1}). Thereafter,
we derive an extended auxiliary system for one of the most simple test cases, the (two-site) Hubbard dimer,
and show how the additionally emerging bath parameters follow from a singular value decomposition of the
offdiagonal components of the self-energy (Sec.~\ref{subsec.subsec2.2.1}). After numerical validation of the
scheme in Sec.~\ref{subsec.subsec2.2.2}, we then generalize the approach to an arbitrary number of sites and
hence to a general quantum system (Sec.~\ref{subsec.subsec2.3}). Based on further numerical results, which are
presented in Sec.~\ref{sec.sec3}, we finally discuss additional simplifications of the method.


\section{Theory}\label{sec.sec2}
For the discussion of the auxiliary Hamiltonian approach and its extension to non-local self-energies, we
consider the single-band Fermi-Hubbard model,
\begin{align}
\label{eq:hubbard_hamiltonian}
H=\sum_{ij\sigma}T_{ij}c_{i\sigma}^\dagger c_{j\sigma}+U\sum_{i}\left(n_{i\uparrow}-\frac{1}{2}\right)\left(n_{i\downarrow}-\frac{1}{2}\right)\;,
\end{align}
where $T_{ij}$ are hopping matrix elements connecting the lattice sites $i$ and $j$, $U$ is the local Coulomb
repulsion, $c_{i\sigma}^\dagger$ and $c_{i\sigma}$ are creation and annihilation operators for particles of
spin $\sigma\in\{\uparrow,\downarrow\}$, and $n_{i\sigma}=c^\dagger_{i\sigma}c_{i\sigma}$ is the density. Using the
nonequilibrium Green's function $G_{ij\sigma}(t,t')=-\ii\langle \tc c_{i\sigma}(t)c^\dagger_{j\sigma}(t')\rangle$,
with $\langle\tc\ldots\rangle=\mathrm{tr}[\tc \exp(S)\ldots]/\mathrm{tr}[\tc \exp(S)]$, action $S=-\ii\intc\dd s\,H(s)$
and contour-ordering operator $\tc$, the time evolution of the system (\ref{eq:hubbard_hamiltonian}) is given
by the Kadanoff-Baym equation
\begin{align}
\label{eq:kbe}
 \sum_k[\ii\pdert\delta_{ik}-T_{ik}]G_{kj\sigma}(t,t')&=\dc(t,t')\delta_{ij}+\sum_k\intc\dd s\,\Sigma_{ik\sigma}(t,s)G_{kj\sigma}(s,t')\,,
\end{align}
which is accompanied by its adjoint equation with $t\leftrightarrow t'$. For simplicity, we incorporate the
Hartree potential $V_i(t)=U(\langle n_{i\bar{\sigma}}(t)\rangle-\frac{1}{2})$ into the hopping matrix, thus
we define as $\Sigma_{ij\sigma}(t,t')$ only the correlation part of the self-energy. As the unit of energy 
(time) we choose the (inverse) hopping $T$.


\begin{figure}
\begin{center}
\includegraphics[width=0.95\textwidth]{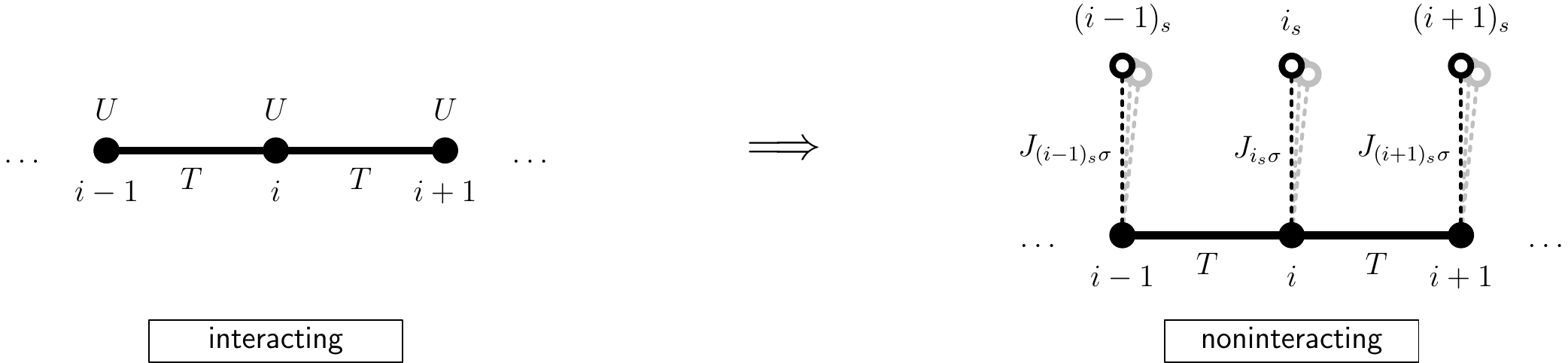} 
\end{center}
\caption{
Construction of the auxiliary Hamiltonian in the case of a local self-energy for a Hubbard chain with
nearest-neighbor hopping $T$ and on-site interaction $U$. While the filled circles denote the original
lattice sites, the open circles mark the bath orbitals $i_s$ which are coupled independently of one
another to a lattice site $i$ by a time- and (in general) spin-dependent hopping matrix element $J_{i_s\sigma}(t)$.
}
\label{fig:fig1}
\end{figure}


\subsection{Auxiliary Hamiltonian for a local self-energy}\label{subsec.subsec2.1}
Following Ref.~\cite{balzer14.aux}, the Kadanoff-Baym equations~(\ref{eq:kbe}) can be mapped onto a 
\textit{noninteracting} auxiliary system when the self-energy is local in space, i.e., if
$\Sigma_{ij\sigma}(t,t')\propto\delta_{ij}\Sigma_{i\sigma}(t,t')$. In this case the retardation effects
described by $\Sigma_{i\sigma}$ can be mimicked by sets ${\cal S}_i$ of bath orbitals which are coupled
to each individual lattice site. The corresponding auxiliary Hamiltonian is given by
\begin{align}
\label{eq:aux_hamiltonian}
H_\aux(t)=&\sum_{ij\sigma}T_{ij}c_{i\sigma}^\dagger c_{j\sigma}+\sum_{i\sigma}\sum_{s}\epsilon_{i_s\sigma}(t)a_{i_s\sigma}^\dagger a_{i_s\sigma}\\
&+\sum_{i\sigma}\sum_{s\in {\cal S}_i}\left(J_{i_s\sigma}(t)a_{i_s\sigma}^\dagger c_{i\sigma} + J_{i_s\sigma}^*(t)c_{i\sigma}^\dagger a_{i_s\sigma}\right)\,,\nonumber
\end{align}
where $a_{i_s\sigma}^\dagger$ and $a_{i_s\sigma}$ describe the creation and annihilation of particles with
spin $\sigma$ in a bath orbital $i_s$ with energy $\epsilon_{i_s\sigma}(t)$ that is assigned to the lattice
site $i$, see  Fig.~\ref{fig:fig1}. By comparing the KBE of the auxiliary system~(\ref{eq:aux_hamiltonian})
and the original lattice system~(\ref{eq:hubbard_hamiltonian}) [Eq.~(\ref{eq:kbe})], one finds that the
auxiliary system has the same Green's function $G_{ij\sigma}(t,t')$ (with $i,j$ being lattice indices)
provided that for all times $t$ and $t'$ on the Keldysh contour the bath parameters can be determined from
\begin{align}
\label{eq:def_selfenergy_local}
 \Sigma_{i\sigma}(t,t')=\sum_{s\in{\cal S}_i}J_{i_s\sigma}(t)g(\epsilon_{i_s\sigma};t,t')J_{i_s\sigma}^*(t')\,,
\end{align}
where $g(\epsilon;t,t')$ is the Green's function of an isolated bath orbital,
\begin{align}
\label{eq:def_isolated_bath gf}
 g(\epsilon;t,t')=\ii[f_\beta(\epsilon(0))-\thc(t,t')]\mathrm{e}^{-\ii\int_{t}^{t'}\dd s\,\epsilon(s)}\,,
\end{align}
with $f_\beta(\epsilon)=1/(\mathrm{e}^{\beta\epsilon}+1)$ being the Fermi-Dirac distribution for an inverse
temperature $\beta$, and $\thc$ the Heavyside step function on the contour.

Analyzing Eq.~(\ref{eq:def_selfenergy_local}) regarding the location of the time arguments on the contour,
one can distinguish two classes ${\cal S}_i^{(1)}\cup{\cal S}_i^{(2)}={\cal S}_i$ of bath orbitals that model
different dynamical effects:
\begin{itemize}
 \item [(i)] For the mixed components ($\Sigma^\rceil_{i\sigma}$) of the self-energy, the bath orbitals in
 ${\cal S}_i^{(1)}$ describe the time evolution of correlations that are present in the initial state. As
 these contributions typically decay as function of time, the corresponding bath orbitals become decoupled
 ($J_{i_s\sigma}(t)\rightarrow0$) in the limit $t\rightarrow\infty$. In general, the bath parameters of this
 class depend on the generalized spectral function 
 \begin{align}
  C^\rceil_{i\sigma}(t,\epsilon)=\frac{1}{2\pi}\left(\Sigma^\rceil_{i\sigma}(t,\epsilon+\ii 0)-\Sigma^\rceil_{i\sigma}(t,\epsilon-\ii 0)\right)\,.
 \end{align}
 Details on the construction of the parameters are given in Ref.~\cite{gramsch13}.
 \item [(ii)] For the real-time greater ($\Sigma^>_{i\sigma}$) and lesser ($\Sigma^<_{i\sigma}$) components
 of the self-energy, the bath orbitals in ${\cal S}_i^{(2)}$ describe the dynamical build-up of correlations.
 If we choose the initial bath energies $\epsilon_{i_s\sigma}(0)$ [at time $t=0$] such that the Fermi-Dirac
 distribution $f_\beta(\epsilon_{i_s\sigma}(0))$ is either $0$ or $1$, the Green's function $g$ evaluates to
 $g^\gtrless(\epsilon_{i_s\sigma};t,t')=\mp\ii$, and one can decompose both real-time components separately,
 \begin{align}
 \label{eq:sigma_decomposition_les}
  -\ii\Sigma_{i\sigma}^<(t,t')&=\sum_{s\in{\cal S}_{i}^<}J_{i_s\sigma}(t)J_{i_s\sigma}^*(t')\,,\\
 \label{eq:sigma_decomposition_gtr}
  \ii\Sigma_{i\sigma}^>(t,t')&=\sum_{s\in{\cal S}_{i}^>}J_{i_s\sigma}(t)J_{i_s\sigma}^*(t')\,,
 \end{align}
 where ${\cal S}_i^<\cup{\cal S}_i^>={\cal S}_i^{(2)}$.
\end{itemize}

The form of Eqs.~(\ref{eq:sigma_decomposition_les}) and (\ref{eq:sigma_decomposition_gtr}) reveals that the
bath parameters $J_{i_s\sigma}(t)$ in ${\cal S}_i^{(2)}$ can be obtained directly from a matrix decomposition
of the two-time self-energy. As the left-hand sides represent Hermitian \textit{and} positive definite matrices
(in $t$ and $t'$), one can either apply an eigenvalue\footnote{In case of an eigenvalue decomposition, the
square roots of the real eigenvalues are integrated into the hopping parameters.} or a Cholesky decomposition,
see~\cite{gramsch13,balzer14.aux} for details. We emphasize that an exact representation requires the number
of bath orbitals to be equal to the number of time steps. To drastically reduce the size of the baths one
however can use low-rank representations, e.g., by performing the sums in Eqs.~(\ref{eq:sigma_decomposition_les})
and~(\ref{eq:sigma_decomposition_gtr}) only over the important eigenvalues or by using a finite number of bath
orbitals to get an exact Cholesky decomposition on the first time steps while fitting the time dependency at
later times~\cite{gramsch13}. Below, we consider only bath orbitals which fall into class (ii), i.e., we start
either from a  noninteracting initial state, a mean-field state or the atomic limit. In all these cases, the
Matsubara and mixed components of the self-energy vanish ($\Sigma_{i\sigma}^\mathrm{M}=\Sigma_{i\sigma}^\lceil=0$),
and consequently the bath orbitals in ${\cal S}_i^{(1)}$ are redundant.

As a result of the mapping procedure, the solution of the interacting many-body problem~(\ref{eq:hubbard_hamiltonian})
has been replaced by the determination of the Green's function $G^\aux_{ij\sigma}(t,t')$ of the \textit{noninteracting}
auxiliary system. The only assumption, at this stage, is the DMFT approximation of a local self-energy. A
self-consistent solution is further prepared by iteration, starting, e.g., from the noninteracting Green's
function. Moreover, we can use the causal property of the Cholesky decomposition (where an increase of the
matrix dimension does not alter the decomposition at earlier times) to get the self-consistent bath parameters
stepwise directly on the time step, see in particular Ref.~\cite{gramsch13}. In practice, we also do not need
to store the Green's function. Instead it is sufficient to save the self-energy which enters the construction
of the bath parameters. For the numerical computation of the auxiliary Green's function, we have developed a
Krylov-based time-propagation scheme (see Appendix of Ref.~\cite{balzer14.aux}), where the greater ($G_{ij\sigma}^>(t_s,t)$)
and lesser component ($[G_{ij\sigma}^<(t,t_s)]^\dagger$) of the Green's function are calculated for all times
$t\leq t_s$ on the time step $t_s$, i.e., on the edge of the lower and upper propagation triangle, compare,
e.g., with Ref.~\cite{stan09}. In addition, OpenMP-based parallelization of the code is achieved by performing
the time stepping for all vectors $\mathbf{G}_{j\sigma}^>(t_s,t)$ with components $G_{ij\sigma}^>(t_s,t)$ [for
all $i$] in parallel, thus for an infinitesimal time step $\delta t$ and times $t\leq t_s$ (and analogously
for the lesser component), we evaluate
\begin{align}
\label{eq:time_propagation}
 \mathbf{G}_{j\sigma}^>(t_s+\delta t,t)=|\mathbf{G}_{j\sigma}^>(t_s,t)|\exp\left[-\ii \mathbf{h}^\aux_\sigma(t)\delta t\right]\frac{\mathbf{G}_{j\sigma}^>(t_s,t)}{|\mathbf{G}_{j\sigma}^>(t_s,t)|}\,,
\end{align}
where $\mathbf{h}^\aux_\sigma(t)$ denotes the one-particle Hamiltonian of the auxiliary system $H_\aux(t)$. We
note that the normalization of the vector $\mathbf{G}_{j\sigma}^>$ on the right-hand side of Eq.~(\ref{eq:time_propagation})
is a precondition to apply the unitary time-evolution operator within the Krylov method~\cite{hochbruck97}.


\subsection{Generalization of the approach to non-local self-energies}\label{subsec.subsec2.2}
From the first point of view, it seems to be not a straightforward endeavor to take into account arbitrary
non-local self-energies in the auxiliary Hamiltonian approach. This is due to the fact that any non-local,
i.e., offdiagonal component of the self-energy does not describe retardation effects on a single site which
is in clear correspondence to a noninteracting site coupled to a bath as discussed in Sec.~\ref{subsec.subsec2.1}.
Moreover, offdiagonal components of $\Sigma_\sigma$ are less symmetric, which raises the question whether it
is at all possible to construct an auxiliary system such that the conservation laws for particle number,
momentum and energy (which come with a conserving approximation) are obeyed. Nevertheless, it is intuitive
to imagine that correlation effects, which originate from non-local parts of the self-energy, should be
representable by additional particle motion that connects different sites of the lattice. In the following,
we show that a valid mapping exists for the Hubbard dimer. How the findings can be generalized to more extended
systems is discussed afterwards (see Sec.~\ref{subsec.subsec2.3}).


\subsubsection{Hubbard dimer}\label{subsec.subsec2.2.1}
\begin{figure}
\begin{center}
\includegraphics[width=0.95\textwidth]{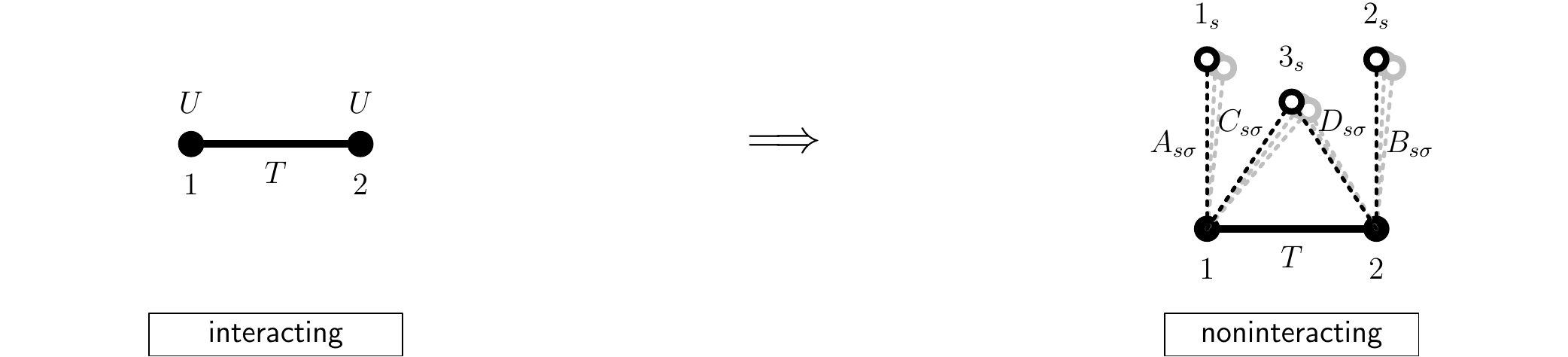} 
\end{center}
\caption{
Construction of the auxiliary Hamiltonian representation of the KBE for the Hubbard dimer, including non-local correlation
effects.
}
\label{fig:fig2}
\end{figure}

The simplest lattice model, in which a non-local self-energy  appears, is the Hubbard dimer~\cite{jafari08,carrascal15}, consisting of two
connected sites $1$ and $2$. In order to account for diagonal \textit{and} offdiagonal components of the self-energy
we extend the auxiliary Hamiltonian of Eq.~(\ref{eq:aux_hamiltonian}) by a set of bath orbitals [with energies
$\epsilon_{3_s\sigma}(t)$] which exchange particles with both sites of the dimer. If we refer to the corresponding
hopping matrix elements as $C_{s\sigma}(t)$ and $D_{s\sigma}(t)$ and define $J_{1_s\sigma}(t)=A_{s\sigma}(t)$
and $J_{2_s\sigma}(t)=B_{s\sigma}(t)$ (compare with Fig.~\ref{fig:fig2}), the ansatz for the mapping is given
by
\begin{align}
\label{eq:aux_hamiltonian_dimer}
 H_\aux(t)=&-T\sum_{\sigma}\left(c_{1\sigma}^\dagger c_{2\sigma}+c_{2\sigma}^\dagger c_{1\sigma}\right)+\sum_{i\in\{1,2,3\}\sigma}\sum_{s\in{\cal S}_i} \epsilon_{i_s\sigma}(t) a_{i_s\sigma}^\dagger a_{i_s\sigma}\\
        &+\sum_{s}\left(A_{s\sigma}(t)a_{1_s\sigma}^\dagger c_{1\sigma}+B_{2_s\sigma}(t)b_{s\sigma}^\dagger c_{2\sigma}+C_{s\sigma}(t)a_{3_s\sigma}^\dagger c_{1\sigma}+D_{s\sigma}(t)a_{3_s\sigma}^\dagger c_{2\sigma}+\mathrm{H.c.}\right)\,,\nonumber
\end{align}
where in the last term $s$ runs only over existing bath sites. To determine the parameters of the baths
$1_s$, $2_s$ and $3_s$ in the presence of all components of the self-energy, we formulate the equations
of motions for the one-particle Green's function of the auxiliary system (\ref{eq:aux_hamiltonian_dimer})
and compare them to the KBE of the Hubbard dimer ($i,j\in\{1,2\}$),
\begin{align}
\label{eq:kbe_dimer}
 [\ii\pdert +T\delta_{|i-j|1}]G_{ij\sigma}(t,t')-\delta_{ij}\dc(t,t')=\sum_{k\in\{1,2\}}\intc\dd s\,\Sigma_{ik\sigma}(t,s)G_{kj\sigma}(s,t')\,.
\end{align}
For the auxiliary system we have
\begin{itemize}
 \item[(i)] the \textit{lattice components} of the auxiliary Green's functions, which evolve in time
 according to
\begin{align}
\label{eq:aux_kbe_dimer1a}
\ii\pdert G^\aux_{1j\sigma}-\delta_{1j}\dc(t,t')&=-TG^\aux_{2j\sigma}(t,t')+\sum_{s}\left[A_{s\sigma}(t)G^\aux_{1_s j\sigma}(t,t')+C_{s\sigma}(t)G^\aux_{3_s j\sigma}(t,t')\right]\,,\\
\ii\pdert G^\aux_{2j\sigma}-\delta_{2j}\dc(t,t')&=-TG^\aux_{1j\sigma}(t,t')+\sum_{s}\left[B_{s\sigma}(t)G^\aux_{2_s j\sigma}(t,t')+D_{s\sigma}(t)G^\aux_{3_s j\sigma}(t,t')\right]\,,\nonumber
\end{align}
where $j\in\{1,2\}$, and
\item[(ii)] the \textit{mixed bath-lattice terms}, which obey
\begin{align}
\label{eq:aux_kbe_dimer1b}
[\ii\pdert -\epsilon_{1_s\sigma}(t)]G^\aux_{1_s j\sigma}&=A^*_{s\sigma}(t)G^\aux_{1j\sigma}(t,t')\,,\\
[\ii\pdert -\epsilon_{2_s\sigma}(t)]G^\aux_{2_s j\sigma}&=B^*_{s\sigma}(t)G^\aux_{2j\sigma}(t,t')\,,\nonumber\\
[\ii\pdert -\epsilon_{3_s\sigma}(t)]G^\aux_{3_s j\sigma}&=\left[C^*_{s\sigma}(t)G^\aux_{1j\sigma}(t,t')+D^*_{s\sigma}(t)G^\aux_{2j\sigma}(t,t')\right]\,.\nonumber
\end{align}
\end{itemize}


Analogously to Ref.~\cite{balzer14.aux},  the mixed bath-lattice terms can be determined by using the
Green's function $g(\epsilon_{i_s\sigma};t,t')$ of an isolated bath orbital (cf.~Eq.~(\ref{eq:def_isolated_bath gf})):
\begin{align}
\label{eq:aux_kbe_dimer2}
 G^\aux_{1_s j\sigma}(t,t')&=\intc\dd s\,g(\epsilon_{1_s\sigma};t,s)A^*_{s\sigma}(s)G^\aux_{1j\sigma}(s,t')\,,\\
 G^\aux_{2_s j\sigma}(t,t')&=\intc\dd s\,g(\epsilon_{2_s\sigma};t,s)B^*_{s\sigma}(s)G^\aux_{2j\sigma}(s,t')\,,\nonumber\\
 G^\aux_{3_s j\sigma}(t,t')&=\intc\dd s\,g(\epsilon_{3_s\sigma};t,s)\left[C^*_{s\sigma}(s)G^\aux_{1j\sigma}(s,t')+D^*_{s\sigma}(s)G^\aux_{2j\sigma}(s,t')\right]\,.\nonumber
\end{align}
Inserting the results of Eq.~(\ref{eq:aux_kbe_dimer2}) into Eq.~(\ref{eq:aux_kbe_dimer1a}) and
comparing term by term to the Kadanoff-Baym equation~(\ref{eq:kbe_dimer}), we find that the
offdiagonal components of the self-energy can indeed be represented by the bath $3_s$. However,
the expressions for the diagonal components do not only involve the hopping matrix elements
$A_{s\sigma}(t)$ and $B_{s\sigma}(t)$. Instead, we obtain
\begin{align}
\label{eq:sigma_dimer}
 \Sigma_{11\sigma}(t,t')&=\sum_{s\in{\cal S}_1}A_{s\sigma}(t)g(\epsilon_{1_s\sigma};t,t')A^*_{s\sigma}(t')+\sum_{s\in{\cal S}_3}C_{s\sigma}(t)g(\epsilon_{3_s\sigma};t,t')C^*_{s\sigma}(t')\,,\\
 \Sigma_{22\sigma}(t,t')&=\sum_{s\in{\cal S}_2}B_{s\sigma}(t)g(\epsilon_{2_s\sigma};t,t')B^*_{s\sigma}(t')+\sum_{s\in{\cal S}_3}D_{s\sigma}(t)g(\epsilon_{3_s\sigma};t,t')D^*_{s\sigma}(t')\,,\nonumber\\
 \Sigma_{12\sigma}(t,t')&=\sum_{s\in{\cal S}_3}C_{s\sigma}(t)g(\epsilon_{3_s\sigma};t,t')D^*_{s\sigma}(t')\,,\nonumber\\
 \Sigma_{21\sigma}(t,t')&=\sum_{s\in{\cal S}_3}D_{s\sigma}(t)g(\epsilon_{3_s\sigma};t,t')C^*_{s\sigma}(t')\,.\nonumber
\end{align}
In passing, we note that the result~(\ref{eq:sigma_dimer}) can also be obtained by integrating
out the bath orbitals in a path integral, which can be done exactly if the bath orbitals appear
only up to quadratic order~\cite{negele98}. Furthermore, the right-hand sides of the last two
lines in Eq.~(\ref{eq:sigma_dimer}) are in line with the Hermitian symmetry of the real-time
quantities, i.e.,
\begin{align}
 \Sigma_{12\sigma}^\gtrless(t,t')=-[\Sigma_{21}^\gtrless(t',t)]^*&\stackrel{\mathrm{def.}}{=}-\sum_{s\in{\cal S}^\gtrless}D^*_{s\sigma}(t')[g^\gtrless(\epsilon_{3_s\sigma};t',t)]^*C_{s\sigma}(t)\\
 &=\sum_{s\in{\cal S}^\gtrless}C_{s\sigma}(t)g^\gtrless(\epsilon_{3_s\sigma};t,t')D^*_{s\sigma}(t')\,,\nonumber
\end{align}
where we used $g^\gtrless(\epsilon;t,t')=-[g^\gtrless(\epsilon;t',t)]^*$ in the last equality.

For the discussion of the result~(\ref{eq:sigma_dimer}) we assume that the initial state of
the Hubbard dimer is not correlated, i.e., $\Sigma^\mathrm{M}=\Sigma^\lceil=0$. To fix the
bath parameters, it is instructive to begin with the decomposition of the offdiagonal components
of the self-energy. This will yield the time-dependent hopping parameters $C_{s\sigma}(t)$
and $D_{s\sigma}(t)$ and, subsequently, allows us to determine the parameters $A_{s\sigma}(t)$
and $B_{s\sigma}(t)$ from the first two lines in Eq.~(\ref{eq:sigma_dimer}). As in Sec.~\ref{subsec.subsec2.1},
we choose the energies of the bath orbitals such that the initial occupations of the bath
orbitals are either $0$ (for the greater component of $\Sigma$) or $1$ (for the lesser
component of $\Sigma$), which leads to 
\begin{align}
 \pm\ii\Sigma_{12\sigma}^\gtrless(t,t')=\sum_{s\in{\cal S}^\gtrless}C_{s\sigma}(t)D^*_{s\sigma}(t')\,.
\end{align}
As function of $t$ and $t'$, the matrices $\ii\Sigma_{12\sigma}^>$  and $-\ii\Sigma_{12\sigma}^<$
are neither Hermitian nor positive definite which generally rules out an eigenvalue and a
Cholesky decomposition. This is in contrast to the diagonal components of the self-energy
discussed in Sec.~\ref{subsec.subsec2.1} (Eqs.~(\ref{eq:sigma_decomposition_les}) and~(\ref{eq:sigma_decomposition_gtr})). 
However, one can perform a singular value decomposition
of the objects as
\begin{align}
\label{eq:svd_dimer}
 \pm\ii\Sigma_{12\sigma}^\gtrless(t,t')=\sum_{s\in{\cal S}^\gtrless_3}U_{s\sigma}(t)S_{s\sigma} V_{s\sigma}(t')\,,
\end{align}
where $U_\sigma$ and $V_\sigma$ denote unitary matrices and $S_{\sigma}$ is a real vector
containing the non-negative singular values. As the singular values $S_{s\sigma}$ typically
decay rapidly in magnitude, we can further use a low-rank decomposition by choosing the number
of bath orbitals smaller than the number of time steps. Incorporating the singular values
symmetrically into the matrices $U_\sigma$ and $V_\sigma$, we obtain the first set of bath
parameters as 
\begin{eqnarray}
\label{eq:svd_result_cd}
 C_{s\sigma}(t)=\sqrt{S_{s\sigma}}U_{s\sigma}(t)\,,\hspace{2.5pc}D_{s\sigma}(t)=\sqrt{S_{s\sigma}}V_{s\sigma}^*(t)\,.
\end{eqnarray}

As a next step we can use the result of Eq.~(\ref{eq:svd_result_cd}) to determine the bath
parameters $A_{s\sigma}(t)$ and $B_{s\sigma}(t)$. Taking into account the initial occupations,
the first two lines of Eq.~(\ref{eq:sigma_dimer}) can be rewritten as ($i=1,2$)
\begin{align}
\label{eq:sigma_dimer_ab1}
\Delta^\gtrless_{i\sigma}(t,t')=\pm\ii\Sigma_{ii\sigma}^\gtrless(t,t')-\Gamma_{i\sigma}^\gtrless(t,t')\,,
\end{align}
with
\begin{align}
\label{eq:sigma_dimer_ab2}
\Delta_{1\sigma}^\gtrless(t,t')&=\sum_{s\in{\cal S}^\gtrless_1}A_{s\sigma}(t)A_{s\sigma}^*(t')\,,&
\Gamma_{1\sigma}^\gtrless(t,t')&=\sum_{s\in{\cal S}^\gtrless_3}C_{s\sigma}(t)C^*_{s\sigma}(t')\,,&\nonumber\\
\Delta_{2\sigma}^\gtrless(t,t')&=\sum_{s\in{\cal S}^\gtrless_2}B_{s\sigma}(t)B_{s\sigma}^*(t')\,,&
\Gamma_{2\sigma}^\gtrless(t,t')&=\sum_{s\in{\cal S}^\gtrless_3}D_{s\sigma}(t)D^*_{s\sigma}(t')\,.
\end{align}
While the matrices $\pm\ii\Sigma^\gtrless_{ii\sigma}$ and $\Gamma^\gtrless_{i\sigma}$ are both
Hermitian and positive definite by construction, the positive definiteness does not hold in
general for their difference $\Delta^\gtrless_{i\sigma}$, Eq.~(\ref{eq:sigma_dimer_ab1}). For
this reason, the partitioning of $\Delta^\gtrless_{i\sigma}$ into the bath parameters $A_{s\sigma}(t)$
and $B_{s\sigma}(t)$, respectively, cannot be performed through a Cholesky decomposition. Instead,
we use an eigenvalue decomposition,
\begin{align}
 \label{eq:evd_dimer}
 \Delta_{1\sigma}^\gtrless(t,t')&=\sum_{s\in{\cal S}^\gtrless_1}P_{s\sigma}(t)E_{s\sigma} P_{s\sigma}^*(t')\,,\\
 \Delta_{2\sigma}^\gtrless(t,t')&=\sum_{s\in{\cal S}^\gtrless_2}Q_{s\sigma}(t)F_{s\sigma} Q_{s\sigma}^*(t')\,,\nonumber
\end{align}
where the eigenvectors form the rows of the matrices $P_\sigma$ and $Q_\sigma$ and the
corresponding eigenvalues are contained in the vectors $E_{\sigma}$ and $F_{\sigma}$.
Comparing Eqs.~(\ref{eq:sigma_dimer_ab2}) and~(\ref{eq:evd_dimer}), we thus obtain the
remaining bath parameters as
\begin{align}
 \label{eq:svd_result_ab}
 A_{s\sigma}(t)=\sqrt{E_{s\sigma}}P_{s\sigma}(t)\,,\hspace{2.5pc}B_{s\sigma}(t)=\sqrt{F_{s\sigma}}Q_{s\sigma}(t)\,.
\end{align}


\begin{figure}
\begin{center}
\includegraphics[height=0.4\textwidth]{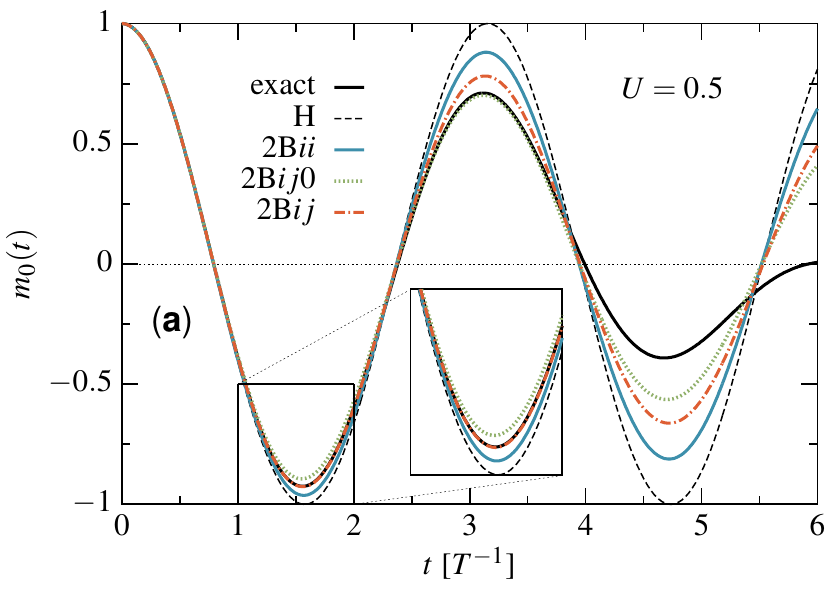}
\includegraphics[height=0.4\textwidth]{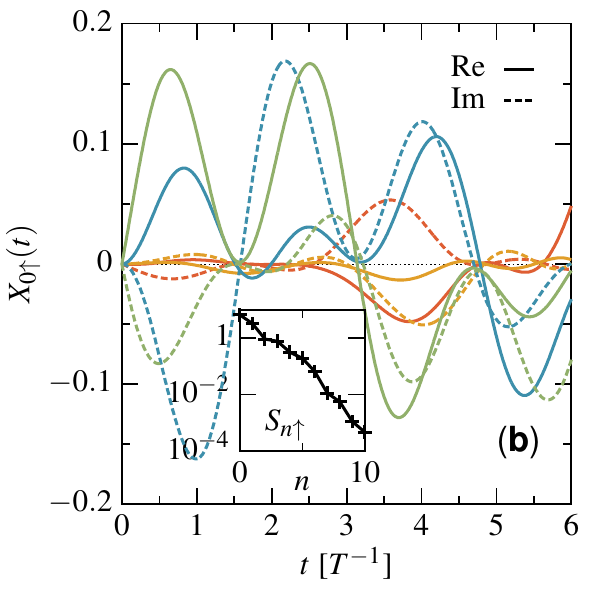}
\end{center}
\caption{
Hubbard dimer prepared in the N\'{e}el state at $t=0$. (a) Time evolution of the staggered
magnetization $m_0(t)=\langle n_{0\uparrow}(t)\rangle-\langle n_{0\downarrow}(t)\rangle$
on the left site of the dimer for $U=0.5$ in Hartree~(H) and second-order Born approximation~(2B),
as computed from the self-consistent auxiliary Green's function $G_{ij\sigma}^\aux(t,t')$. The
line labeled 2B$ij$ (2B$ii$) denotes the result for a non-local (local) self-energy. The curve
denoted as 2B$ij0$ shows the result where $\Gamma_{i\sigma}$ is additionally set to zero in
Eq.~(\ref{eq:sigma_dimer_ab1}). (b)~Time-dependent hopping parameters $A_{0\sigma}(t)$ (red
lines), $B_{0\sigma}(t)$ (orange lines), $C_{0\sigma}(t)$ (blue lines) and $D_{0\sigma}(t)$
(green lines). The inset shows the singular values of the decomposition~(\ref{eq:svd_dimer}) of the non-local
self-energy $\Sigma_{12\sigma}(t,t')$.
}
\label{fig:fig3}
\end{figure}


\subsubsection{Numerical validation}\label{subsec.subsec2.2.2}
The derivation presented in the previous section proofs that there exists a procedure
which maps the full Kadanoff-Baym equations of the Hubbard dimer onto a noninteracting
auxiliary model. In this section, we demonstrate the validity of the mapping also numerically.
To this end, we prepare the Hubbard dimer (at time $t=0$) in the perfect N\'{e}el state
$|\Psi_0\rangle=c_{2\downarrow}^\dagger c_{1\uparrow}^\dagger|0\rangle$ and use the
second-order Born approximation of the self-energy,
\begin{align}
\label{eq:second_born_se}
 \Sigma_{ij\sigma}^{\mathrm{2B},\gtrless}(t,t')=U^2G_{ji\sigma}^\gtrless(t,t') G_{ji\bar{\sigma}}^\gtrless(t,t') G_{ij\bar{\sigma}}^\lessgtr(t',t)\,,
\end{align}
to monitor the decay of the local staggered magnetization $m_i(t)=\langle n_{i\uparrow}(t)\rangle-\langle n_{i\downarrow}(t)\rangle$
in the weak-coupling regime. For convenience, we compute the Green's function only for one
spin direction. The opposite component follows from the symmetry $G_{ij\bar{\sigma}}(t,t')=G_{(L+1-i)(L+1-j)\sigma}(t,t')$,
where $L=2$ is the spatial dimension of the dimer.

Fig.~\ref{fig:fig3}a shows the self-consistent second Born result for an on-site interaction
$U=0.5$, where the KBE has been solved with a time step of size $\delta t=0.01$ (on a time
window $[0,6]$) including $n_\aux=20$ orbitals for each bath~$i_s$. The result is further
compared to the Hartree dynamics (given by the black dashed line) and to the exact dynamics
(see the black solid line), which has been obtained by exact diagonalization. On the short
time scale, unlike the mean-field result, correlations induce a damping of the oscillatory
magnetization dynamics in the dimer. If we consider nothing but the local parts of the self-energy,
i.e., if we set to zero the bath parameters $C_{s\sigma}(t)$ and $D_{s\sigma}(t)$ in the
auxiliary system (Eq.~(\ref{eq:aux_hamiltonian_dimer})), the time evolution of the magnetization
follows the blue solid line (labeled 2B$ii$). The inclusion of the non-local self-energy
improves this result as expected, see the red dash-dotted line referred to as 2B$ij$. Particularly,
we find good agreement with the exact dynamics for times $t\lesssim2.5$. As additional result
(denoted 2B$ij0$), we present the green dotted curve, where in Eq.~(\ref{eq:sigma_dimer_ab2})
the quantities $\Gamma_{1\sigma}^\gtrless$ and $\Gamma_{2\sigma}^\gtrless$ have been neglected
in the determination of the bath parameters $A_{s\sigma}(t)$ and $B_{s\sigma}(t)$. We note that
such an incomplete solution also leads to an acceptable dynamics, although there are discrepancies
at short times. Moreover, we emphasize that all schemes conserve the particle number as function
of time.

To further illustrate the method, Fig.~\ref{fig:fig3}b shows the time dependence of the
complex hopping parameters $A_{0\sigma}(t)$ to $D_{0\sigma}(t)$, i.e., for the first bath
orbital $s=0$ in each set ${\cal S}^<_i$. Starting from zero (at time $t=0$ where the self-energy
$\Sigma_{ij\sigma}^\gtrless$ vanishes), the time evolution of the bath parameters is rather
complicated despite the simplicity of the Hubbard dimer. Moreover, the inset in Fig.~\ref{fig:fig3}b
displays the singular values $S_{n\uparrow}$ of Eq.~(\ref{eq:svd_dimer}) which decay exponentially
in magnitude and thus allow one to use a low-rank representation of the bath $3_s$. Concerning the
baths $1_s$ and $2_s$, we have included the $20$ most important eigenvalues of each self-energy
decomposition, i.e., those which have the largest absolute values. Note that an exact representation
of the self-energy using a time step of $\delta=0.01$ would require the dimension of each individual
bath to be $n_\aux=600$ on the considered time interval.


\subsection{Generalization to an arbitrary lattice system}\label{subsec.subsec2.3}
\begin{figure}
\begin{center}
\includegraphics[width=0.95\textwidth]{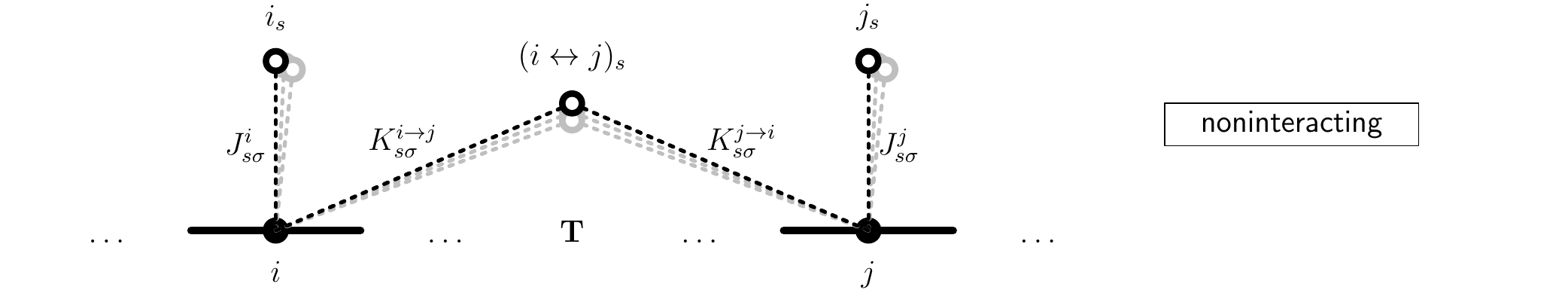} 
\end{center}
\caption{
Generalization of the auxiliary Hamiltonian approach to an extended lattice system with a hopping
matrix $\mathbf{T}$ and an arbitrary non-local self-energy $\Sigma_{ij\sigma}^\gtrless(t,t')$.
Together with the dashed lines, the open circles show the layout of the bath and how it is connected
to the lattice sites (filled circles) by complex hopping matrix elements $K_{s\sigma}^{i\rightarrow j}(t)$,
$K_{s\sigma}^{j\rightarrow i}(t)$, $J_{s\sigma}^{i}(t)$ and $J_{s\sigma}^{j}(t)$.
}
\label{fig:fig4}
\end{figure}

For a lattice system with more than two sites, the self-energy should have a similar auxiliary
representation as discussed above in Sec.~\ref{subsec.subsec2.2.1}. Without giving the proof, we
state that the \textit{offdiagonal} components of the self-energy can always be expressed in the
form ($i\neq j$)
\begin{align}
\label{eq:general_case_offdiagonal_se}
 \pm\ii\Sigma_{ij\sigma}^\gtrless=\sum_{s\in{\cal S}^\gtrless_{i\leftrightarrow j}}K_{s\sigma}^{i\rightarrow j}(t)[K_{s\sigma}^{j\rightarrow i}(t')]^*\,,
\end{align}
with time-dependent hopping parameters $K^{i\rightarrow j}_{s\sigma}(t)$ connecting a lattice site $i$
and a bath orbital~$s$. In the course of this, it is important to remark that the parameters $K^{i\rightarrow j}_{s\sigma}(t)$
are different from $K^{j\rightarrow i}_{s\sigma}(t)$. However, both sets follow from a single singular
value decomposition of the left-hand side of Eq.~(\ref{eq:general_case_offdiagonal_se}). On the other
hand, the auxiliary representation for the \textit{diagonal} components of the self-energy is given by
\begin{align}
\label{eq:general_case_diagonal_se}
 \pm\Sigma_{ii\sigma}^\gtrless=\sum_{s\in{\cal S}^\gtrless_i}J_{s\sigma}^{i}(t)[J_{s\sigma}^{i}(t')]^*+\sum_{j\neq i}\sum_{s\in{\cal S}^\gtrless_{i\leftrightarrow j}}K_{s\sigma}^{i\rightarrow j}(t)[K_{s\sigma}^{i\rightarrow j}(t')]^*\,,
\end{align}
where the bath parameters $J_{s\sigma}^{i}(t)\equiv J_{i_s\sigma}(t)$ are defined as in Sec.~\ref{subsec.subsec2.1}
and can be obtained from an eigenvalue decomposition applied to Eq.~(\ref{eq:general_case_diagonal_se}).
For an illustration of the resulting auxiliary system see Fig.~\ref{fig:fig4}.


In summary, the one-particle auxiliary Hamiltonian which replaces the full Kadanoff-Baym equation
of Eq.~(\ref{eq:kbe}) for an arbitrary non-local self-energy is of the form
\begin{align}
\label{eq:general_case_1pauxham} 
\mathbf{h}^\aux_\sigma(t)=
\left(
\begin{array}{ccc}
 \mathbf{T} & \mathbf{J}_\sigma & \mathbf{K}_\sigma\\
 \mathbf{J}^\dagger_\sigma & \mathbf{0} & \mathbf{0}\\
 \mathbf{K}^\dagger_\sigma & \mathbf{0} & \mathbf{0}
\end{array}
\right)\,,
\end{align}
where the matrix elements of $\mathbf{T}$ are the hopping parameters $T_{ij}$ of the original Hubbard
model~(\ref{eq:hubbard_hamiltonian}), the matrix $\mathbf{J}$ is diagonal involving the bath parameters
$J_{s\sigma}^i(t)$, and the matrix $\mathbf{K}$ includes the bath parameters $K^{i\rightarrow j}_{s\sigma}(t)$
and $K^{j\rightarrow i}_{s\sigma}(t)$. If we use $n_\aux$ bath orbitals for the representation of each
component $\Sigma_{ij\sigma}$ (with $i\leq j$) of the self-energy and consider a Hubbard model with $L$ sites, the dimension of the auxiliary
Hamiltonian $\mathbf{h}^\aux_\sigma$ is given by 
\begin{align}
 {\cal D}_\aux=L+2 n_\aux L+n_\aux (L^2-L)=(n_\aux+1)L+L^2\,,
\end{align}
i.e., the size of the generalized auxiliary system scales linearly with the size of the bath and
quadratically with the number of lattice sites. Moreover, we emphasize that the low-rank representation
of Eqs.~(\ref{eq:general_case_offdiagonal_se}) and~(\ref{eq:general_case_diagonal_se}) [with fixed
$n_\aux\ll n_t$] generically leads to a time propagation scheme which scales quadratically with the
number of time steps $n_t$. This is one of the main advantages of the method and allows the scheme
to outperform standard KBE solvers.


\section{Application of the method}\label{sec.sec3}
To validate the generalization of the auxiliary Hamiltonian approach for a quantum system which is more
complex than the Hubbard dimer discussed in Sec.~\ref{subsec.subsec2.2.2}, we now consider a Hubbard chain
with $L=10$ sites. For this system it is still easy to compute the dynamics exactly, and therefore it is
possible to judge upon the quality of the second-order Born approximation, cf.~Eq.~(\ref{eq:second_born_se}).
Similarly to the analysis performed on exact grounds in Ref.~\cite{bauer15}, we study the time evolution of the N\'{e}el
state, i.e., the initial state is given by the one-particle density matrix
\begin{align}
\rho_{ij\sigma}(0)=-\ii G^<_{ij\sigma}(0,0)=-\ii\delta_{ij}\langle\Psi_0|n_{i\sigma}|\Psi_0\rangle\,,
\end{align}
where
\begin{align}
 |\Psi_0\rangle=\prod_{i\in A}c_{i\downarrow}^\dagger\prod_{j\in B}c_{j\uparrow}^\dagger|0\rangle\,,
\end{align}
and $A$ and $B$ denote the two sublattices. To perform the simulations, we choose $n_\aux=30$, $\delta t=0.01$
and iterate until self-consistency is reached on the time interval $[0,6]$.

\begin{figure}
\begin{center}
\includegraphics[height=0.325\textwidth]{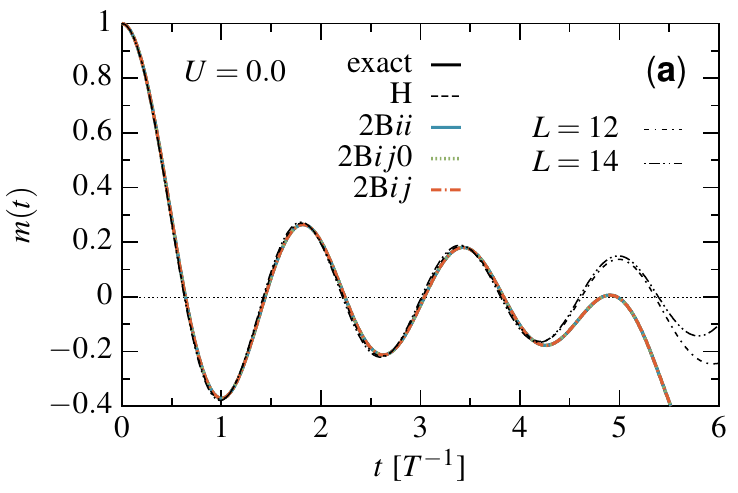}
\includegraphics[height=0.325\textwidth]{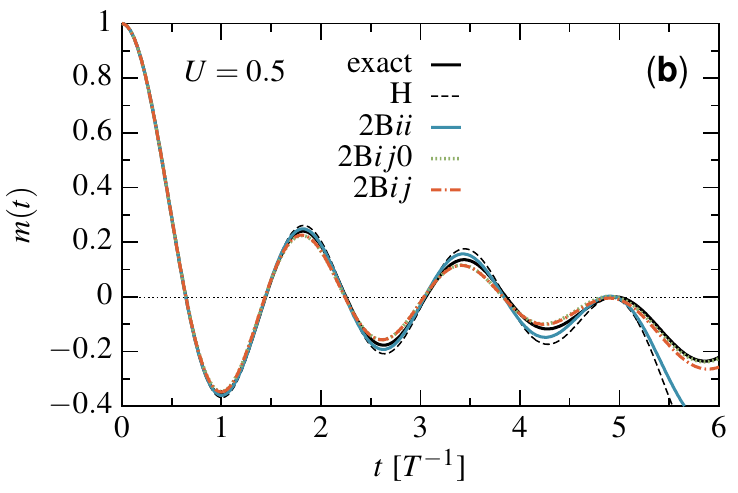}
\includegraphics[height=0.325\textwidth]{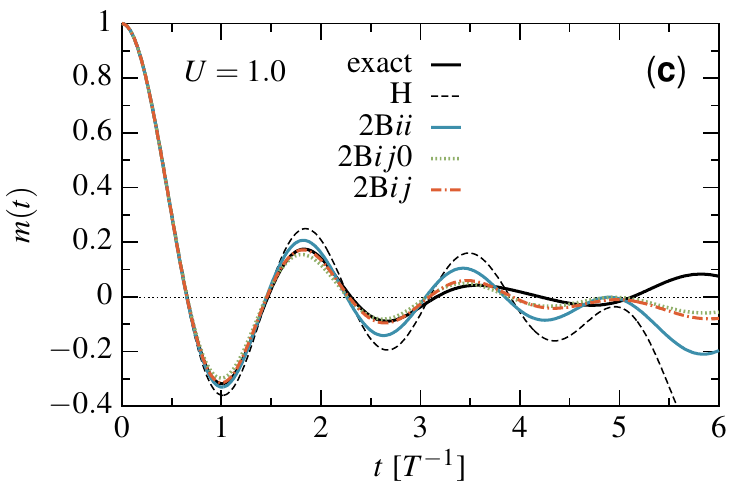}
\includegraphics[height=0.325\textwidth]{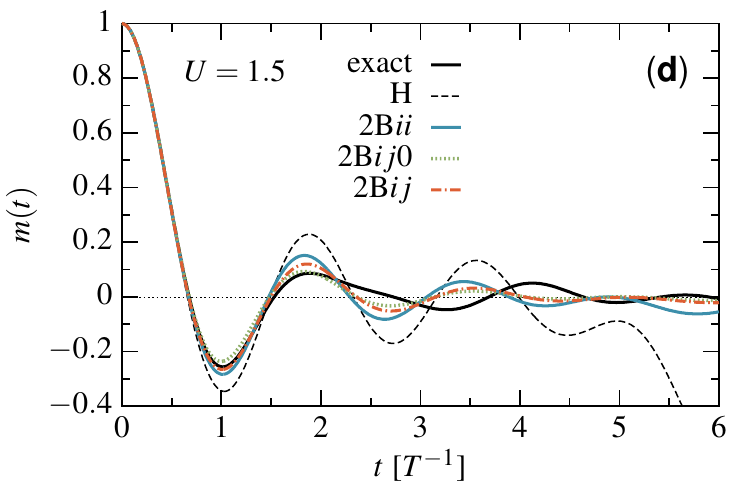}
\includegraphics[height=0.325\textwidth]{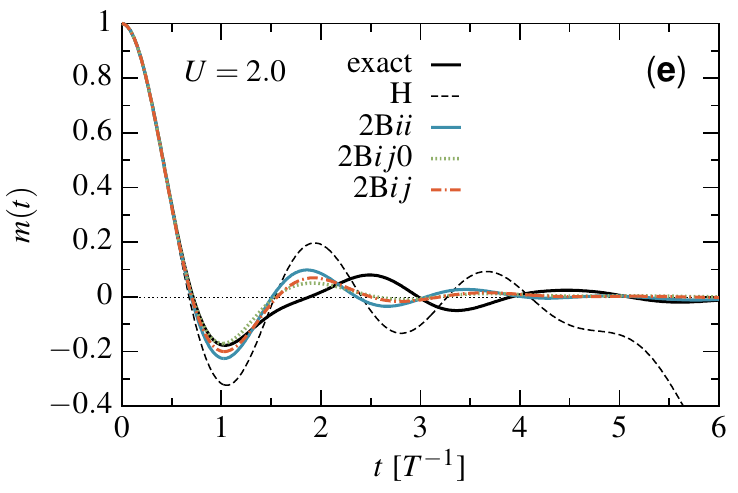}
\includegraphics[height=0.325\textwidth]{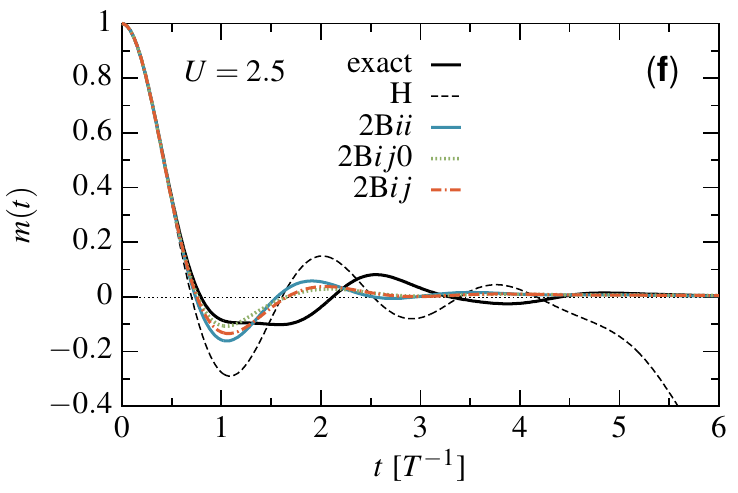}
\includegraphics[height=0.325\textwidth]{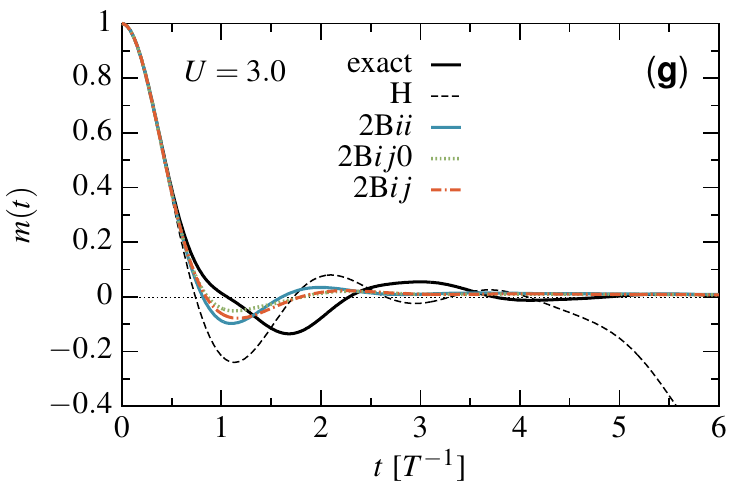}
\includegraphics[height=0.325\textwidth]{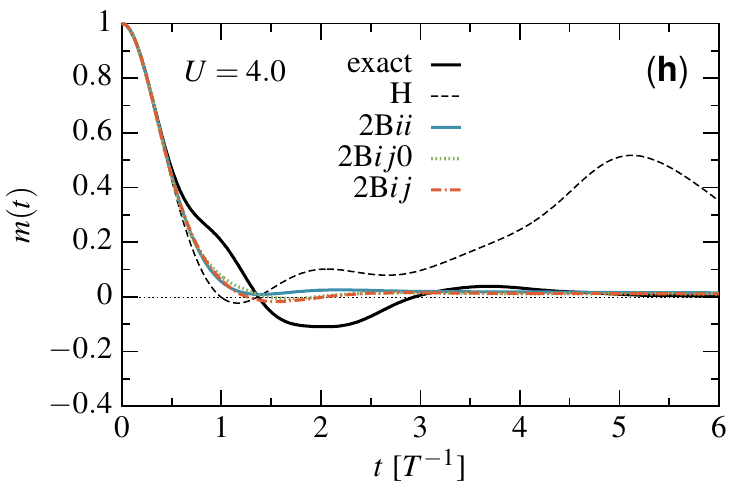}
\end{center}
\caption{(a)-(h): Self-consistent results for the decay of the magnetization $m(t)$ in an open $10$-site Hubbard
chain with nearest-neighbor hopping $T$ and on-site interaction $U$, which ranges from zero to $4.0$. While
the black dashed (solid) line shows the Hartree (exact) result, the colored lines correspond to different
levels of the second-order Born approximation, recall Fig.~\ref{fig:fig3} in Sec.~\ref{subsec.subsec2.2.2}. In 
panel~(a), we also include exact results for a chain with $L=12$ and $14$ sites.
}
\label{fig:fig5}
\end{figure}

We consider two cases, first a Hubbard chain with open boundary conditions and second a closed chain,
respectively, a ring. Figure.~\ref{fig:fig5} summarizes the results for the former system, showing the
time evolution of the average magnetization $m(t)=\tfrac{2}{L}\sum_{i\in A}(\langle n_{i\uparrow}(t)\rangle-\langle n_{i\downarrow}(t)\rangle)$
for different approximation levels of the self-energy and different strengths of the interaction $U$.
Generally, we find that the antiferromagnetic order parameter melts on the time scale of a few hopping
times at weak interaction. In the limit $U\rightarrow 0$ and large $L$ (see Fig.~\ref{fig:fig5}a), we
observe a damped oscillation of $m(t)$ which is of the form of a zeroth-order Bessel function $J_0(\gamma t)$
with some constant $\gamma$, compare, e.g., with Ref.~\cite{barmettler09}. At small $U\lesssim1$, the
second-order Born approximation well describes the relaxation
of the magnetization, whereas the Hartree approximation (H) leads to larger oscillations of $m(t)$. These
deficiencies of the mean-field results particularly increase for larger interactions. On the contrary,
the second Born approximation (also for larger $U$) correctly captures the time scale on which the order
parameter relaxes towards $m=0$, see, e.g., Fig.~\ref{fig:fig5}f. Furthermore, the inclusion of
non-local parts of the second-order Born self-energy (see the lines labeled 2B$ij$ in comparison to
2B$ii$) leads to results which agree better with the exact data. This is of course expected because a
DMFT approximation generally performs inadequately in low-dimensional systems.

\begin{figure}
\begin{center}
\includegraphics[height=0.325\textwidth]{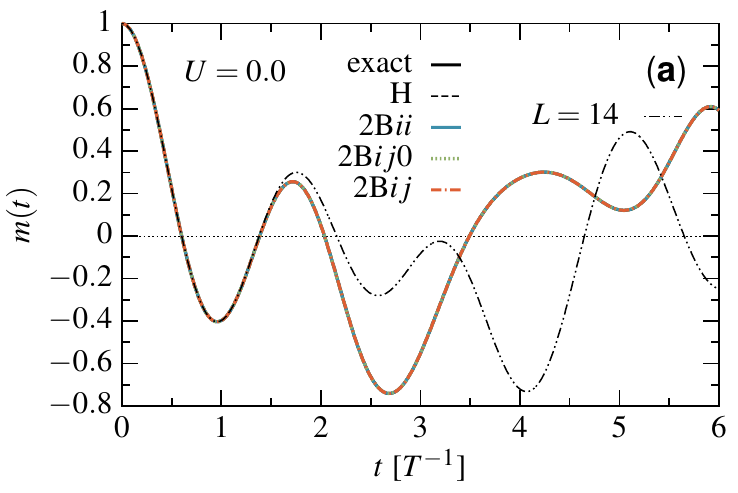}
\includegraphics[height=0.325\textwidth]{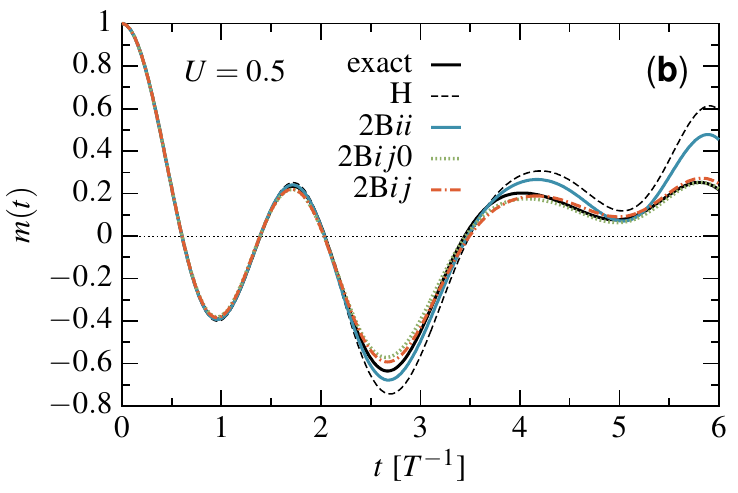}
\includegraphics[height=0.325\textwidth]{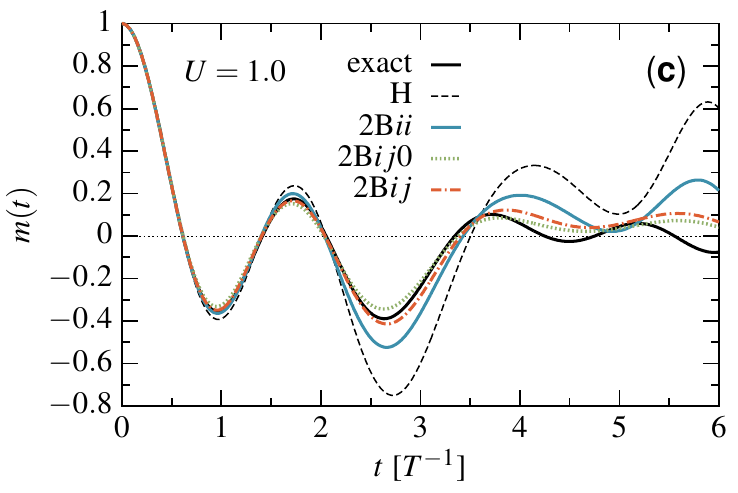}
\includegraphics[height=0.325\textwidth]{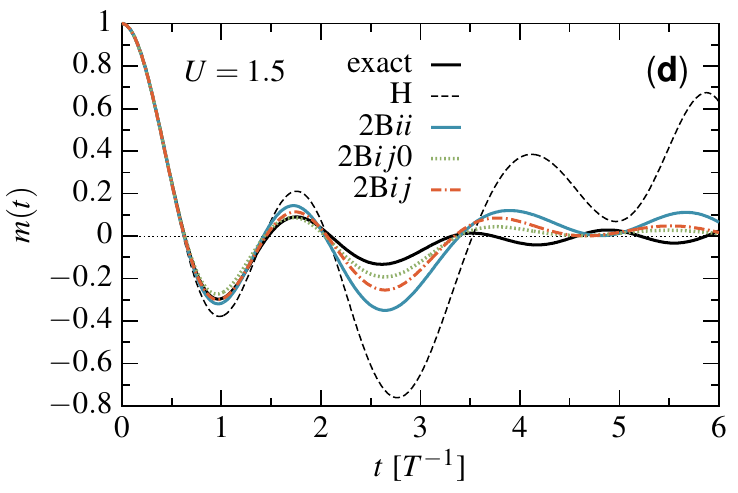}
\includegraphics[height=0.325\textwidth]{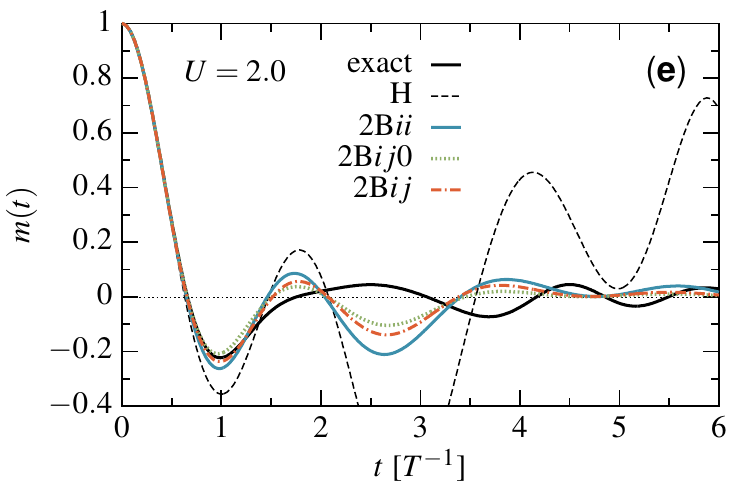}
\includegraphics[height=0.325\textwidth]{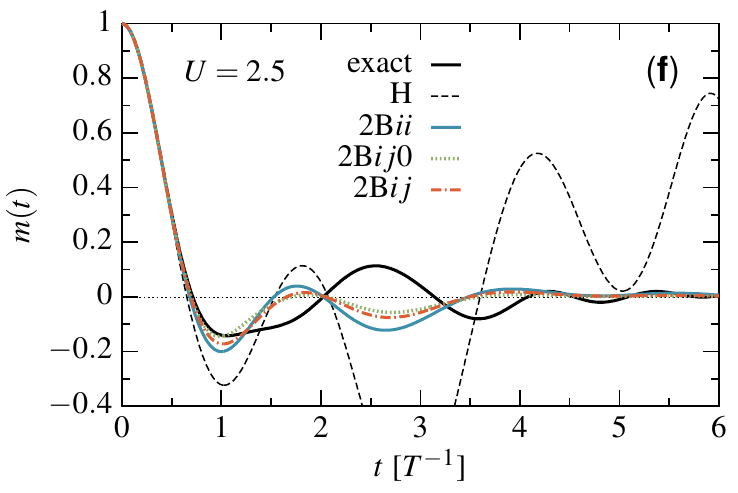}
\includegraphics[height=0.325\textwidth]{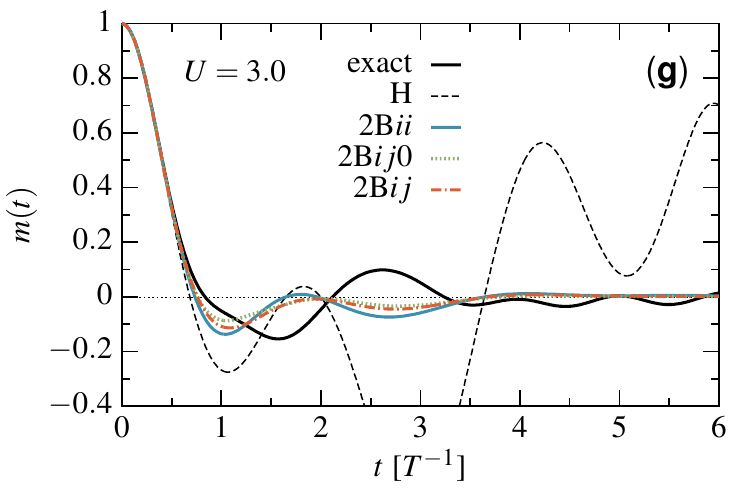}
\includegraphics[height=0.325\textwidth]{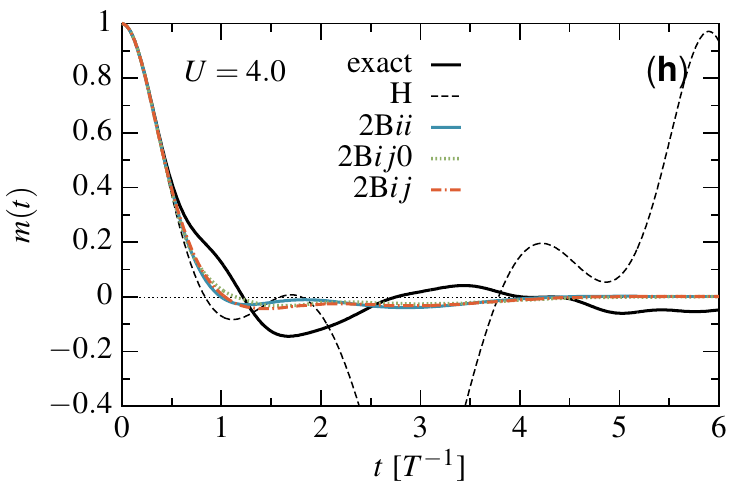}
\end{center}
\caption{
Same as in Fig.~\ref{fig:fig5} but for a closed Hubbard chain, i.e., a ring with $10$ sites where fermions
can additionally hop between sites $1$ and $10$.
}
\label{fig:fig6}
\end{figure}

In Fig.~\ref{fig:fig6} we show the simulations for the closed chain. Here, the oscillations of the Hartree
results are even more dominant for larger values of the interaction ($U\gtrsim1.5$), and correlations
are essential to describe the dynamics of the magnetization. It is not our intention, in this paper, to
discuss the physics behind the melting process. The time evolution is basically driven by the fact that
local magnetic moments decay along with the long-range order at sufficiently small interactions, whereas
in the Mott insulating regime (large $U$) energy is transferred from charge excitations to the spin background
while local moments persist~\cite{balzer15.neel}. By means of nonequilibrium DMFT simulations for the
Bethe lattice it has further been shown in Ref.~\cite{balzer15.neel} that the dynamics at weak interaction
is governed by residual quasiparticles, which generate oscillations in the offdiagonal components of the
momentum distribution. 


In both types of systems, we finally observe that the conservation of particles, which is an essential test
for the numerical implementation, is not destroyed when the bath parameters $K^{i\rightarrow j}_{s\sigma}(t)$
and $K^{j\rightarrow i}_{s\sigma}(t)$ are \textit{not} taken into account in the decomposition of the diagonal parts
of the self-energy (Eq.~(\ref{eq:general_case_diagonal_se})). Instead, the corresponding time evolution of
the magnetization (see the green dotted lines) is relatively close to the full second-order Born approximation
where these terms are included, compare with the red dash-dotted lines. We can therefore identify another
time propagation scheme where the diagonal and offdiagonal parts of the self-energy are decomposed independently.
This means an additional simplification of the method, because the associated matrices become Hermitian and
positive definite again. However, we note that one still looses the causal structure in the time propagation
as a singular value decomposition is required to treat Eq.~(\ref{eq:general_case_offdiagonal_se}).


\section{Conclusion}\label{sec.sec4}
The intriguing possibility to map \textit{any} KBE with a given self-energy onto a noninteracting auxiliary
system means that there exists a general alternative scheme to solve two-time quantum kinetic equations.
This scheme does not at all require the evaluation of the memory kernel and can bypass the $n_t^3$-scaling by
representing the two-time self-energies in terms of their most important ``modes''. These modes appear
as time-dependent couplings to bath orbitals which are connected to either one or two (lattice) coordinates
and thus mimic the retardation effects of the interacting system which are described by the local and non-local
parts of the self-energy.
In the course of this,
all the properties which come with the self-energy are retained by the
scheme, i.e., in particular, a $\Phi$-derivable approximation automatically leads to an auxiliary system in
which particle number, energy and momentum are preserved.
In general, the Hamiltonian representation can exploit its full potential when the calculation of the self-energy
scales essentially better than $n_t^3$ which is the case for the second Born approximation. However, the method
should be efficiently applicable also for some more advanced self-energies, e.g., for the GW approximation, where
a similar two-time decomposition of the polarization could be used to self-consistently determine the screened
interaction.


\section*{Acknowledgements}\label{sec.sec4}
We thank E.~Arigoni and C.~Verdozzi for stimulating discussions and acknowledge M.~Eckstein for
carefully reading the manuscript.




\section*{References}
\begin{thebibliography}{9}

\bibitem{kadanoff62} Kadanoff L P and Baym G 1962 Quantum Statistical Mechanics (W.A. Benjamin, Inc. New York)

\bibitem{keldysh64} Keldysh L V 1964 \textit{Zh.~Eksp.~Teor.~Fiz.}~\textbf{47} 1515 [1965 \textit{Sov.~Phys.~JETP} \textbf{20} 1018]

\bibitem{stefanucci13.cup} Stefanucci G and van Leeuwen R 2013 Nonequilibrium Many-Body Theory of Quantum Systems: A Modern Introduction (Cambridge University Press, Cambridge)

\bibitem{danielewicz84.a} Danielewicz P 1984 \textit{Annals of Phys.}~\textbf{152} 239

\bibitem{danielewicz84.b} Danielewicz P 1984 \textit{Annals of Phys.}~\textbf{152} 305

\bibitem{kohler01} K\"{o}hler H S and Morawetz K 2001 \textit{Phys.~Rev.~C} \textbf{64} 024613

\bibitem{kwong00} Kwong N-H and Bonitz M 2000 \textit{Phys.~Rev.~Lett.}~\textbf{84} 1768

\bibitem{kremp05} Kremp D, Schlanges M, and Kraeft W-D 2005 Quantum Statistics of Nonideal Plasmas (Springer, Berlin)

\bibitem{kwong98} Kwong N-H,  Bonitz M,  Binder R, and K\"{o}hler H S 1998 \textit{phys.~stat.~sol.~(b)} \textbf{206} 197

\bibitem{gartner99} Gartner P,  B\'{a}nyai L, and Haug H 1999 \textit{Phys.~Rev.~B} \textbf{60} 14234

\bibitem{lorke06}Lorke M, Nielsen T R, Seebeck J, Gartner P, and Jahnke F 2006 \textit{Phys.~Rev.~B} \textbf{73} 085324

\bibitem{freericks06}Freericks J K,  Turkowski V M, and Zlati\'{c} V 2006 \textit{Phys.~Rev.~Lett.}~\textbf{97} 266408

\bibitem{aoki14} Aoki H, Tsuji N, Eckstein M, Kollar M, Oka T, and Werner P 2014 \textit{Rev.~Mod.~Phys.}~\textbf{86} 779

\bibitem{dahlen07} Dahlen N E and van Leeuwen R 2007 \textit{Phys.~Rev.~Lett.}~\textbf{98} 153004

\bibitem{perfetto15} Perfetto E, Uimonen A-M, van Leeuwen R, and Stefanucci G 2015 \textit{Phys.~Rev.~A} \textbf{92} 033419

\bibitem{morawetz99} Morawetz K and K\"{o}hler H S 1999 \textit{Eur.~Phys.~J.~A} \textbf{4} 291

\bibitem{kremp00} Kremp D, Bornath T, Bonitz M,  Kraeft W D, and Schlanges M 2000 \textit{Phys.~Plasmas} \textbf{7} 59

\bibitem{kohler99.fortran} K\"{o}hler H S, Kwong N H, and Yousif H A 1999 \textit{Comp.~Phys.~Comm.}~\textbf{123} 123

\bibitem{rios11} Rios A, Barker B, Buchler M, and Danielewicz P 2011 \textit{Annals of Phys.}~\textbf{326} 1274

\bibitem{hermanns12.gkba} Hermanns S, Balzer K, and Bonitz M 2012 \textit{Physica Scripta} \textbf{T151} 014036

\bibitem{semkat99} Semkat D, Kremp D, and Bonitz M 1999 \textit{Phys.~Rev.~E} \textbf{59} 1557

\bibitem{morozov99} Morozov V G and  R\"{o}pke G 1999 \textit{Annals of Phys.}~\textbf{278} 127

\bibitem{dahlen05} Dahlen N E and van Leeuwen R 2005 \textit{J.~Chem.~Phys.}~\textbf{122} 164102

\bibitem{stan09} Stan A, Dahlen N E, and van Leeuwen R 2009 \textit{J.~Chem.~Phys.}~\textbf{130} 224101

\bibitem{balzer10.pra1} Balzer K, Bauch S, and Bonitz M 2010 \textit{Phys.~Rev.~A} \textbf{81} 022510

\bibitem{garny10} Garny M and M\"{u}ller M M 2010 High Performance Computing in Science and Engineering, Garching/Munich 2009 (Springer, Berlin)

\bibitem{balzer10.pra2} Balzer K, Bauch S, and Bonitz M 2010 \textit{Phys.~Rev.~A} \textbf{82} 033427

\bibitem{balzer13.lnp} Balzer K and Bonitz M 2013 \textit{Lect.~Notes Phys.}~\textbf{867}

\bibitem{lipavsky86} Lipavsk\'y  P, \ifmmode \check{S}\else \v{S}\fi{}pi\ifmmode \check{c}\else \v{c}\fi{}ka V, and Velick\'y B 1986 \textit{Phys.~Rev.~B} \textbf{34} 6933

\bibitem{hermanns13.jpcs} Hermanns S, Balzer K, and Bonitz M 2013 \textit{J.~Phys.:~Conf.~Ser.}~\textbf{427} 012008

\bibitem{latini14} Latini S, Perfetto E, Uimonen A-M, van Leeuwen R, and Stefanucci G 2014 \textit{Phys.~Rev.~B} \textbf{89} 075306

\bibitem{balzer13.jpcs} Balzer K, Hermanns S, and Bonitz M 2013 \textit{J.~Phys.:~Conf.~Ser.}~\textbf{427} 012006

\bibitem{hermanns14.prb} Hermanns S, Schl\"{u}nzen N, and Bonitz M 2014 \textit{Phys.~Rev.~B} \textbf{90} 125111

\bibitem{freericks08} Freericks J K 2008 \textit{Phys.~Rev.~B} \textbf{77} 075109

\bibitem{gramsch13} Gramsch C, Balzer K, Eckstein M, and Kollar M 2013 \textit{Phys.~Rev.~B} \textbf{88} 235106

\bibitem{wolf14.mps} Wolf F A, McCulloch I P, and Schollw\"{o}ck U 2014 \textit{Phys.~Rev.~B} \textbf{90} 235131

\bibitem{balzer15.mctdh} Balzer K, Li Z, Vendrell O, and Eckstein M 2015 \textit{Phys.~Rev.~B} \textbf{91} 045136

\bibitem{balzer15.neel} Balzer K, Wolf F A, McCulloch I P, Werner P, and Eckstein M 2015 \textit{Phys.~Rev.~X}~\textbf{5} 031039

\bibitem{balzer14.aux} Balzer K and Eckstein M 2014 \textit{Phys.~Rev.~B} \textbf{89} 035148

\bibitem{gramsch15} Gramsch C and Potthoff M 2015 \textit{Phys.~Rev.~B} \textbf{92} 235135

\bibitem{hochbruck97}Hochbruck M and Lubich C 1997 \textit{SIAM J.~Numer.~Anal.}~\textbf{34} 1911

\bibitem{jafari08} Jafari S A 2008 \textit{Iranian J.~Phys.~Res.}~\textbf{8} 113

\bibitem{carrascal15} Carrascal D J, Ferrer J, Smith J C, and Burke K 2015 \textit{J.~Phys.:~Cond. Mat.}~\textbf{27} 393001

\bibitem{negele98} Negele J W and Orland H 1998 Quantum Many-Particle Systems (Westview Press, Boulder, CO)

\bibitem{bauer15} Bauer A, Dorfner F, and Heidrich-Meisner F 2015 \textit{Phys.~Rev.~A} \textbf{91} 053628

\bibitem{barmettler09} Barmettler P, Punk M, Gritsev V, Demler E, and Altman E 2009 \textit{Phys.~Rev.~Lett.}~\textbf{102} 130603

\end{thebibliography}
\end{document}